\documentstyle[psfig,12pt]{article}
\def\reference{\parskip 0pt\par\noindent\hangindent 0.5 truecm}
\def\kms{km ${\rm s}^{-1}$}

\newcommand{\HI}{H\thinspace\protect\footnotesize I\protect\normalsize}
\newcommand{\HIb}{H\thinspace\protect\large\bf I\protect\Large \bf}
\newcommand{\hi}{H\thinspace{\protect\scriptsize I}}

\newcommand{\ij}{\mbox{$I-J$}}
\newcommand{\jk}{\mbox{$J-K$}}
\newcommand{\B}{{$B$}}

\newcommand{\II}{{$I_c$}}
\newcommand{\J}{{$J$}}
\newcommand{\K}{{$K_s$}}
\newcommand{\tfr}{Tully\,--\,Fisher relation}

\newcommand{\etal}{{\it et~al.}}
\newcommand{\cf}{{\it cf.\,}}
\newcommand{\eg}{{\it e.g.},\ }         
\newcommand{\ie}{{\it i.e.},\ }         

\def\la{\mathrel{\hbox{\rlap{\hbox{\lower4pt\hbox{$\sim$}}}\hbox{$<$}}}}
\def\ga{\mathrel{\hbox{\rlap{\hbox{\lower4pt\hbox{$\sim$}}}\hbox{$>$}}}}

\def\deg{{^\circ}}
\def\arcmin{\hbox{$^\prime$}}
\def\arcsec{\hbox{$^{\prime\prime}$}}

\def\fm{\hbox{$.\!\!^{\rm m}$}}

\def\fdg{\hbox{$.\!\!^\circ$}}
\def\farcm{\hbox{$.\mkern-4mu^\prime$}}

\begin{document}
%
%
\title{DENIS Observations of Multibeam Galaxies in the Zone of Avoidance}
%


\author{A. Schr\"oder $^{1,2}$ \and
R.C. Kraan-Korteweg $^{3}$ \and
G.A. Mamon $^{4,5}$ 
} 

\date{}
\maketitle

{\center
$^1$ Australia Telescope National Facility, CSIRO, PO Box 76, Epping, NSW
2121, Australia \\ aschroed@atnf.csiro.au \\[3mm]
$^2$ Institute of Astronomy, NCU, Chung-Li, 32054 Taiwan \\ [3mm]
$^3$ Depto. de Astronomia, Univ. de Guanajuato, Ap. P. 144,
Guanajuato, GTO
36000, Mexico \\ kraan@norma.astro.ugto.mx \\[3mm]
$^4$ IAP, 98 bis Blvd Arago, 75014 Paris, France \\ gam@iap.fr \\[3mm]
$^5$ DAEC, Observatoire de Paris-Meudon, 92195 Meudon, France \\ [3mm]
}

%
\begin{abstract}

Roughly
25\% of the optical extragalactic sky is obscured by the dust and stars of
our Milky Way. Dynamically important structures might still lie hidden in this
zone. Various surveys are presently being employed to uncover the galaxy
distribution in the Zone of Avoidance (ZOA) but all suffer from (different)
limitations and selection effects.

We illustrate the promise of using a multi-wavelength approach for
extragalactic large-scale studies behind the ZOA, \ie a combination of
three surveys -- optical, systematic blind \hi\ and near-infrared
(NIR), which will allow the mapping of the peculiar velocity field in
the ZOA through the NIR \tfr .  In particular, we present here the
results of cross-identifying \hi -detected galaxies with the DENIS NIR
survey, and the use of NIR colours to determine foreground
extinctions.

\end{abstract}

{\bf Keywords: galaxies: distances -- galaxies: photometry -- large scale
structure of the universe }
\bigskip
%
%

\section{Introduction }

Understanding the origin of the peculiar velocity of the Local Group and the
dipole in the Cosmic Microwave Background is one of the major goals of the study
of large-scale structures. Theoretical reconstruction methods, however, still
suffer from large interpolation uncertainties across the Zone of Avoidance
(ZOA), which extends over about 25\% of the optically visible extragalactic sky.
Dynamically im\-portant structures might still lie hidden in this zone, such as
the recently discovered nearby Dwingeloo galaxy (Kraan-Korteweg \etal\ 1994)
and the rich massive cluster Abell~3627 (Kraan-Korteweg \etal\ 1996).  Important
large-scale structures, \eg the Supergalactic Plane and other filaments and
wall-like structures, seem to continue across this zone. Results from various
theoretical approaches still vary strongly, however, in predicting clusters and
voids even in the nearby universe (Sigad \etal\ 1998). This is mainly due to
differences in the theoretical reconstruction methods and the adopted
parameters, the different galaxy samples to which they are applied, and 
the selected interpolation mechanisms to bridge the scarcity of data in the
ZOA. A more complete knowledge of the distribution of galaxies in redshift space
as well as in distance space will improve the reconstructed galaxy density fields and
reduce the diversity in the theoretical predictions.

Various approaches are presently being employed to uncover the galaxy
distribution in the ZOA (\cf Kraan-Korteweg \& Woudt, this volume): deep optical
searches, NIR surveys (DENIS and 2MASS), far-infrared (FIR) surveys (\eg IRAS),
and blind \HI\ searches.  All methods produce new results, but all suffer from
limitations and selection effects.  The combination of data from an optical
galaxy search, a NIR survey and a systematic blind \HI\ survey will allow us to
examine the large-scale structures behind the southern Milky Way and the
peculiar velocity field associated with them.  Redshift independent distance
estimates can be obtained via the NIR \tfr .

A deep optical survey in the southern ZOA is being conducted by one of
us (Kraan-Korteweg \&~Woudt 1994, Kraan-Korteweg \etal\ 1995, 1996,
and references therein). In this region ($265\deg \la \ell \la
340\deg$, $|b| \la 10\deg$), over 11\,000 previously unknown galaxies
above a diameter limit of $D\!=\!0\farcm2$ and with \mbox{$B \la
19\fm0-19\fm5$} have been identified, next to the previously known
$\sim\!300$ Lauberts galaxies with $D\!=\!1^{\prime}$ (Lauberts
1982). Galaxies can be identified visually for \mbox{$A_B \la
5^{\rm m}$} (or typically down to about $|b| \simeq 5\deg$ depending on
surface brightness and Galactic longitude).

The Multibeam (MB) survey of the ZOA, a systematic blind \HI\ survey
(Staveley-Smith 1997), will penetrate even the deepest extinction
layers and trace gas-rich spirals out to redshifts of 12\,000\,\kms
. The survey is currently being conducted with the Multibeam Receiver
(13 beams in the focal plane array) at the 64\,m Parkes telescope and
will cover the whole southern ZOA ($213\deg \la \ell \la 33\deg$; $|b|
\la 5\deg$) with a $5 \sigma$ detection limit of 10\,mJy.  Hardly any of the
1500 predicted detections are expected to have
an optical counterpart, but at these latitudes many might be
visible in the NIR. This new prospect becomes feasible with the
currently ongoing NIR surveys such as DENIS (DEep NIR southern sky
Survey; Epchtein 1997, Epchtein \etal\ 1997) and 2MASS (Skrutskie
\etal\ 1997).  First results from the DENIS survey, which is expected
to cover the entire southern sky by the year 2001, are discussed in
the following sections.

\section{The DENIS survey}

Observations in the NIR can provide important complementary data to other surveys. NIR
surveys will\\
\indent $\bullet$ be sensitive to early-type galaxies --- tracers of massive
groups and clusters --- which are missed in IRAS and \HI\ surveys,\\
\indent $\bullet$ have less confusion with Galactic objects compared to FIR
surveys,\\
\indent $\bullet$ be less affected by absorption than optical surveys.\\
Furthermore, because recent star formation contributes only little to the flux
at this wavelength, the NIR gives a better estimation of the stellar mass
content of galaxies and is hence ideally suited for the application of the \tfr .  

The DENIS survey has currently imaged 40\% of the southern sky in the
$I\,(0.8\,\mu)$, $J\,(1.25\,\mu)$ and $K_s (2.15\,\mu)$ passbands with
a resolution of $1\arcsec$ in \II\ and $3\arcsec$ in \J\ and \K .
In a pilot study, we have assessed the performance of the DENIS survey
at low Galactic latitudes (Schr\"oder \etal\ 1997, hereafter Paper I,
Kraan-Korteweg \etal\ 1998, hereafter Paper II). We (a) tested the
potential of the DENIS survey to detect galaxies where optical and FIR
surveys fail, \ie at high foreground extinctions and in crowded
regions, (b) established that the NIR colours of galaxies lead to
values for the foreground extinction, (c) determined preliminary
$I_c^o$, $J^o$ and $K_s^o$ galaxy luminosity functions in the rich
cluster A3627, and (d) cross-identified highly obscured galaxies
detected in a blind \HI\ search at $|b| < 5\deg$.

\begin{figure}[tb]
\vspace{-2cm}
\begin{center}
\hfil \psfig{file=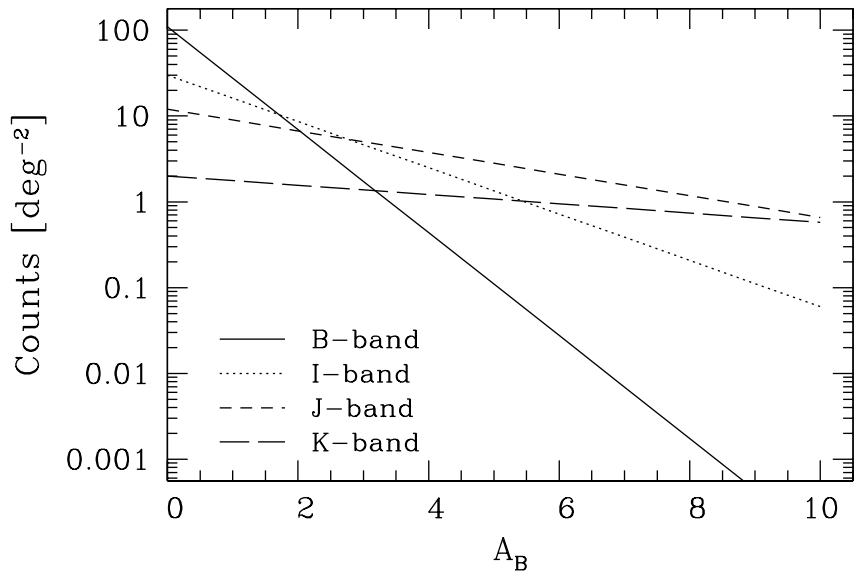,height=10cm}\hfil  
\caption{Predicted galaxy counts in \B , \II , \J\ and \K\ as a function of
absorption in \B , for highly complete and reliable DENIS galaxy samples and a
$B_J \leq 19^{\rm m}$ optical sample.  }
\label{galctsplot}           
\end{center}
\end{figure}

Number counts of galaxies are lower in the ZOA due to the high foreground
extinction, but the effect depends on wavelength. Interpolating from Cardelli
\etal\ (1989), the extinctions in the NIR passbands are $A_{I_c}\!=\!0\fm45$,
$A_J\!=\!0\fm21$, and $A_{K_s}\!=\!0\fm09$ for $A_B\!=\!1\fm0$, hence the
decrease in number counts as a function of extinction is considerably slower in
the NIR than in the optical. Figure~\ref{galctsplot} shows the predicted surface
number density of galaxies for DENIS (Mamon et al. 1998) and for $B < 19$, as a
function of Galactic foreground extinction (\cf Paper II).

The NIR becomes notably more efficient at $A_B \ga 2-3^{\rm m}$. The \J -band is
the most efficient passband to find galaxies at intermediate extinctions, but at
$A_B \simeq 10^{\rm m}$ \K\ becomes superior to \J . Even at $A_B \!>\!  10^{\rm
m}$ we can expect to find galaxies in \J\ and \K . A new air conditioning system
for the \K -band camera installed in April 1997 has decreased its instrumental background by
$0\fm7$, which makes the \K -band now more comparable to \J . These are very rough
predictions and do not take into account any dependence on morphological type,
surface brightness, orientation and crowding, which may lower the number of
actually detectable galaxies (Mamon 1994).

\section{Cross-identification of \HIb  -detected galaxies on DENIS images }

Figure~\ref{mbmapplot} displays the distribution of MB galaxies in the ZOA
(large circles). Dots indicate galaxies found in the optical B-band search with
a diameter limit of $D = 0\farcm2$ (Woudt \etal\ 1998).  Contours indicate
extinction levels determined from the DIRBE maps (Schlegel \etal\ 1998), which
are scaled to independent extinction determinations from Mg$_2$-indices in this
region (Woudt 1998).  The thick contour corresponds to $A_B = 2\fm75$; this is
the completeness limit for galaxies with an extinction-corrected diameter of
$D^{o}\!=1\farcm3$ in the deep optical ZOA galaxy catalogue. The other contours
indicate $A_B=5^{\rm m}$ and $10^{\rm m}$, where the Milky Way is opaque in the
optical.

\begin{figure}[tb]\vspace{-2cm}
\begin{center}
\vspace{13cm}
\caption{Galaxy distribution in part of the southern ZOA with galaxies found in an optical search
(small dots) and galaxies detected with the shallow MB survey (large circles). The
superimposed contours represent absorption levels of $A_B = 2\fm75$ ({\it thick
line}), $5\fm0$, $10\fm0$ (see text for details). MB galaxies for which DENIS
data are available are marked with a large dot, those re-identified on DENIS
images with a cross, and those without an optical counterpart with a star.  }
\label{mbmapplot}           
\end{center}
\end{figure}

For 24 of the galaxies detected in the shallow MB-ZOA survey (\cf\ Henning
\etal , this volume), DENIS survey images ($12\arcmin \times
12\arcmin$) covering the full positional uncertainty region
($\sim\!4\arcmin \times \sim\!4\arcmin$) were currently
available. They are displayed as large dots (including 8 detections
for which only partial coverage of the \HI\ positional uncertainty
region was available on existing DENIS images).  For 16 out of the 24
a clear counterpart could be identified (crosses); of these, 6 are
invisible in the B-band due to large foreground extinctions
(stars). For 4 out of the 24 MB-ZOA galaxies the cross-identifications
are uncertain, and for the remaining 4 galaxies no counterparts could
be identified. These galaxies either lie behind a very thick
extinction layer (\eg one galaxy at $b \simeq 0\deg$ with $A_B \sim
70^{\rm m}$), or they are late-type spirals or irregular galaxies of
very low surface brightness and hence below the magnitude limits of
the DENIS survey. The \HI\ survey, however, is not affected by the
foreground extinction, and spiral galaxies can be found in the NIR at
lower Galactic latitudes and higher foreground extinction levels than
in the optical.

In Figure~\ref{mb5plot}, DENIS images of five MB galaxies are presented. The \II
-band is shown in the upper panel, the \J -band in the middle and the \K -band
in the bottom panel. From left to right the foreground extinction increases.
The first galaxy is a nearby ($v\!=\!1445\,$\kms ) Lauberts galaxy (ESO223-G12)
at $b=+4\fdg8$ and $A_B = 3\fm9$. It is prominent in all three NIR passbands
(note the larger image scale for this galaxy, \ie $3\farcm3$ instead of
$1\farcm7$).  The second galaxy, also detected with IRAS, at $(b,A_B) =
(-3\fdg4,4\fm4$) is slightly more distant ($v\!=\!2789\,$\kms ). This galaxy has
also been identified in \B\ and is quite distinct in the NIR.  The third and
fourth galaxies at $(b,A_B) \simeq (3\fdg4, 4\fm8)$ and $(2\fdg3, 8\fm4)$,
respectively, have no optical counterparts. Here, the extinction effects are
quite obvious: while both have a similar appearance in the \J -band, the third
is very faint in \K\ and the fourth very faint in \II .  The fifth example has
been detected at the highest extinction so far: $(b,A_B,v) \simeq
(+1\fdg2,11\fm8, 3963\,$\kms).  It is fully obscured in the \II -band and barely
visible in the \J -band.

\begin{figure}[tb]
\begin{center}
\vspace{12cm}
\caption{DENIS survey images (before bad pixel filtering) of five galaxies
detected in \hi\ with the MB survey at $|b| \le 5\deg$; the \II\ band image is
at the top, \J\ in the middle, and \K\ at the bottom. }
\label{mb5plot}           
\end{center}
\end{figure}

Cross-identifications of galaxies detected in the MB survey are not always
unambiguous. The positional accuracy for galaxies detected in the blind \HI\
search is of the order of $\sim\!4\arcmin$. Sometimes more than one possible
counterpart lies within this area as illustrated in Figure~\ref{ambigplot}. The
position of this MB galaxy is indicated by a white rectangle on the DENIS \II
-band image; the foreground extinction is $A_B=4.0$.  Four galaxies can be seen
in its vicinity: an edge-on spiral with a small companion (right and above the
rectangle), an early-type spiral (left) and a late-type spiral with low surface
brightness (left and above the early-type spiral). Although the last seems the
most likely candidate given the \HI\ parameters ($v\!=\!2903\,$\kms ,
$w\!=\!177\,$\kms , $I\!=\!13.1\,$Jy\,\kms) and the other galaxies may be
further away or have less \HI\ mass and hence not be detectable with the shallow
\HI\ survey, the cross-identifications are not always clearcut.  For such cases,
follow-up observations, either with \HI\ mapping or optical and NIR
spectroscopy, are required.

\begin{figure}[tb]
\vspace{-1cm}
\begin{center}
\vspace{10cm}
\caption{A $12\arcmin \times 12\arcmin$ DENIS \II -band image of an MB
detection 
(white rectangle) and four galaxy candidates (with an uncertainty radius of $\sim 4\arcmin$) .  }
\label{ambigplot}           
\end{center}
\end{figure}

\section{Photometry }

We have used a preliminary galaxy pipeline (Mamon \etal\ 1997b, 1998), based
upon the SExtractor package (Bertin \& Arnouts 1996) to obtain \II , \J\ and \K\
Kron photometry from the DENIS images at the location of MB galaxies.  Although
many of the galaxies have a considerable number of stars superimposed on their
images, magnitudes derived from this automated algorithm agree well with
independent measurements.

\begin{figure}[tb]
\begin{center}
\hfil\psfig{file=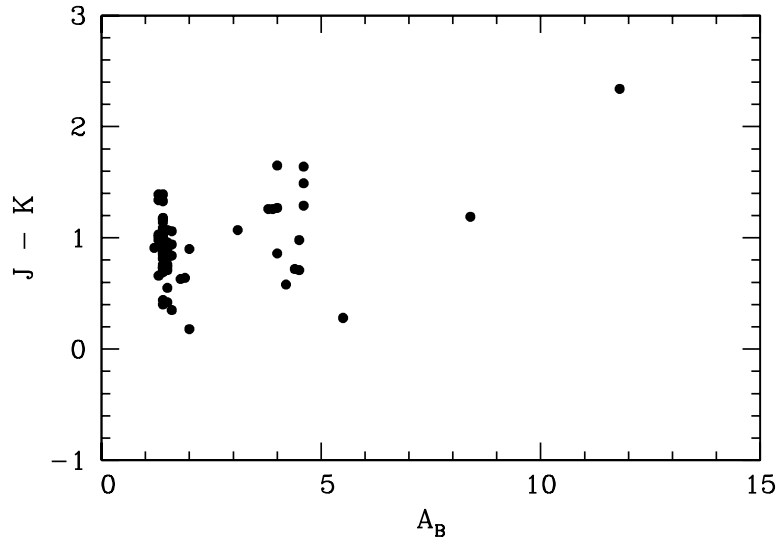,height=10cm} \vspace {-2cm}\hfil
\caption{NIR colour as a dependence on foreground extinction. }
\label{mbextplot}           
\end{center}
\end{figure}

Figure~\ref{mbextplot} shows the dependence of the colour \jk\ on foreground
extinction $A_B$. Included are data from the low-latitude ($\ell\!=\!325\deg$,
$b\!=\!-7\deg$), rich cluster Abell 3627 (\cf\ Paper II), where the foreground
extinction is less than $2^{\rm m}$. The group of MB galaxies with $2^{\rm
m}<A_B<5^{\rm m}$ shows a similar scatter but is offset in colour with
respect to the galaxies in A3627. Also, the galaxy at $A_B=11\fm8$ is
considerably redder than all the other galaxies.

In Paper I we discussed the NIR colour\,--\,colour diagram \ij\
versus \jk\ of galaxies in the A3627 cluster in detail. We found that
the extinction-corrected colours of the cluster galaxies do match the
colours of galaxies in unobscured high latitude regions (Mamon \etal\
1997b, 1998), suggesting that our preliminary photometry is reasonably
accurate.  The shift in colour can be fully explained by the
foreground extinction. Hence, the NIR colours of obscured galaxies (in
particular of elliptical galaxies) provide an independent way of
mapping the extinction in the ZOA (see also Mamon \etal\ 1997a) and
calibrating the DIRBE maps at low Galactic latitudes (see also Mamon
\etal\ 1997a).

\section{Conclusions and future plans}

A combination of the deep optical, systematic blind \HI\ and NIR surveys
illustrates the potential of a multi-wavelength approach for extragalactic
large-scale studies behind the Milky Way. It furthermore allows the mapping of
the peculiar velocity field across the ZOA through the NIR \tfr.

At intermediate latitudes and extinction ($5\deg\!\!<\!\!|b|\!\!<\!\!10\deg$,
$1^{\rm m}\!\!\la\!A_B\!\la\!4-5^{\rm m}$), optical surveys remain superior for
identifying galaxies due to their fainter magnitude limits.  However, the NIR
luminosities and colours will
prove invaluable in analysing the optical survey data and their distribution in
redshift space, and in the final merging of these data with existing sky
surveys.  Despite the high extinction and the star crowding at these latitudes,
\II , \J\ and \K\ photometry from the survey data can be successfully performed
and lead, for instance, to preliminary $I_c^o$, $J^o$ and $K_s^o$ galaxy
luminosity functions in A3627 (Paper II).

At low latitudes and high extinction ($|b| < 5\deg$ and $A_B \ga 4-5^{\rm m}$),
the search for `invisible' obscured galaxies on existing DENIS images shows that
NIR surveys can trace galaxies down to about $|b| \simeq 1\fdg5$.  The \J -band
was found to be optimal for identifying galaxies up to $A_B \simeq 7^{\rm
m}$. NIR surveys can hence further reduce the width of the ZOA. They are
furthermore the only tool that permits the mapping of early-type galaxies ---
tracers of density peaks --- at high extinction.

The blind \HI\ survey uncovers spiral galaxies independent of foreground
extinction. For a significant fraction, a DENIS counterpart has been
found. These MB-ZOA data covers the Galactic latitude range $|b| <\!5\deg$. We
will complement this area with pointed \HI\ observations of optically identified
spiral galaxies for intermediate latitudes ($5\deg\!< |b| <\!10\deg$).
About 300 spiral galaxies have already been detected (Kraan-Korteweg
\etal\ 1997).

In the near future, we plan on obtaining deep \K -band follow-up observations of
the deep and complete \HI -survey in the southern ZOA (\cf Juraszek, this
volume). This will allow us (1) to study the effects of extinction on the extent
and magnitudes of disks in the NIR, (2) to obtain a representative sample of
galaxies across the ZOA for the application of the \K -band \tfr , and (3) to
study the morphology of the detected galaxies.  The last will also be useful to
understand the nature of the large-scale structures revealed with the blind \HI\
survey, in particular as this galaxy sample can readily be merged with the
southern sky \HI\ survey (Kilbourne \etal , this volume).






\section*{Acknowledgements}

We thank Jean Borsenberger for providing bias subtracted, flat fielded DENIS
images, Emmanuel Bertin for supplying recent updates of his SExtractor software
package, and Eric Copet for providing software to display Figures~\ref{mb5plot}
and~\ref{ambigplot}. We also thank Sebastian Juraszek, Elaine Sadler and
Patricia Henning for supplying us with their results of the MB survey of the
ZOA.


\section*{References}






\reference Bertin, E., Arnouts, S. 1996, A\&AS 117, 398



\reference Cardelli J.A., Clayton G.C., Mathis J.S. 1989, ApJ 345, 245

\reference Epchtein, N. 1997,
in {\it The Impact of Large Scale Near-Infrared Surveys} p. 15, eds.\
F. Garzon et al., Kluwer, Dordrecht

\reference Epchtein, N., \etal\ 1997, Messenger 87, 27


\reference Henning, P.A., Staveley-Smith, L., Kraan-Korteweg, R.C., Sadler,
E. 1999, PASA, this volume

\reference Juraszek, S. 1999, PASA, this volume 

\reference Kilbourne, V.A., Webster, R.L., Staveley-Smith, L. 1999, PASA, this volume


\reference Kraan-Korteweg, R.C., Woudt, P.A. 1994, in {\it Unveiling Large-Scale
  Structures Behind the Milky Way}, p. 89, eds.\ C. Balkowski,
  R.C. Kraan-Korteweg, ASP Conf. Ser. 67


\reference Kraan-Korteweg R.C., Loan A.J., Burton W.B. \etal\ 1994, Nature 372, 77

\reference Kraan-Korteweg, R.C., Fairall, A.P., Balkowski, C. 1995,
  A\&A 297, 617

\reference Kraan-Korteweg R.C., Woudt P.A., Cayatte V. \etal\ 1996, Nature 379, 519

\reference Kraan-Korteweg, R.C., Woudt, P.A., Henning, P.A. 1997,
  PASA 14, 15

\reference Kraan-Korteweg, R.C., Schr\"oder, A., Mamon, G., Ruphy, S. 1998,
  in 3rd Euroconference on {\it The Impact of Near-Infrared Surveys on
   Galactic and Extragalactic Astronomy}, p. 209, ed.\ N. Epchtein, Kluwer, Dordrecht 
   (Paper II)


\reference Kraan-Korteweg, R.C., Woudt, P.A. 1999, PASA, this volume

\reference Lauberts, A. 1982,
  The ESO/Uppsala Survey of the ESO (B) Atlas, ESO, Garching
  ESO/Uppsala Survey of the ESO (B) Atlas, ESO, Garching

\reference Mamon G.A. 1994, in
  {\it Unveiling Large-Scale Structures Behind the Milky Way}, p. 53, eds.\
  C. Balkowski, R.C. Kraan-Korteweg, ASP Conf. Ser. 67 

\reference Mamon, G.A., Banchet, V., Tricottet, M. Katz, D. 1997a,
  in {\it The Impact of Large-Scale Near-Infrared Surveys}, p. 239, eds.\
  F. Garzon et al., Kluwer, Dordrecht

\reference Mamon, G.A., Tricottet, M., Bonin, W., Banchet, V. 1997b,
  in XVIIth Moriond Astrophysics Meeting on
  {\it Extragalactic Astronomy in the Infrared}, p. 369, eds.\ G. A. Mamon, 
  Trinh  Xu\^an Thu\^an,  and J. Tr\^an Thanh V\^an,  Fronti\`eres, Paris

\reference Mamon, G.A., Borsenberger, J., Tricottet, M., Banchet, V. 1998,
  in 3rd Euroconference on {\it The Impact of Near-Infrared Surveys on
  Galactic and Extragalactic Astronomy}, p. 177, ed.\ N. Epchtein, Kluwer,
  Dordrecht



\reference Schlegel, D.J., Finkbeiner, D.P., Davis, M. 1998, ApJ 500, 525

\reference Schr\"oder, A., Kraan-Korteweg, R.C., Mamon, G.A. Ruphy, S. 1997,
  in XVIIth Moriond Astrophysics Meeting on
  {\it Extragalactic Astronomy in the Infrared}, p. 381,
  eds. G. A. Mamon, Trinh  Xu\^an  Thu\^an,  and J. Tr\^an Thanh V\^an,
  Fronti\`eres, Paris (Paper I)

\reference Sigad, Y., Eldar, A. Dekel, A. \etal\ 1998, ApJ 495, 516

\reference Staveley-Smith, L. 1997, PASA 14, 111

\reference Skrutskie, M.F., \etal\ 1997,
in {\it The Impact of Large Scale Near-Infrared Surveys}, p. 25,
   eds.\ F. Garzon et al., Kluwer, Dordrecht

\reference Woudt, P.A., Kraan-Korteweg, R.C., Fairall, A.P. \etal\ 1998, A\&A, 338, 8

\reference Woudt, P.A., 1998, Ph.D. thesis, Univ. of Cape Town


\end{document}